# Spatially Resolved Conductivity of Rectangular Interconnects Considering Surface Scattering- Part II: Circuit-Compatible Modeling

Xinkang Chen, and Sumeet Kumar Gupta, Senior *Member,* IEEE

***Abstract*—In Part I of this work, we had presented a spatially resolved model for conductivity of interconnects capturing surface scattering based on the well-known Fuchs-Sondheimer (FS) approach. However, the proposed spatially resolved FS (SRFS) model involves computing complicated integrals making it ill-suited for circuit simulations. In this part, we build upon our SRFS model to develop a circuit-compatible conductivity model for rectangular interconnects accounting for 2D surface scattering. The proposed circuit-compatible model offers spatial resolution of conductivity as well as explicit dependence on the physical parameters such as electron mean free path ($\lambda_0$), specularity ($p$) and interconnect geometry. We validate our circuit-compatible model over a range of physical parameters showing a close match with the physical SRFS model proposed in Part I (with error < 0.7%). We also compare our circuit-compatible model with a previous spatially resolved analytical model (appropriately modified for a fair comparison) and show that our model captures the spatial resolution of conductivity and the dependence on physical parameters more accurately. Finally, we present a semi-analytical equation for the average conductivity based on our circuit-compatible model.**

## I. INTRODUCTION

The progression of technology scaling stands as a key driver for enhancing the electronic devices, facilitating improvements in speed, energy efficiency, and integration density. As we venture into sub-7nm technology nodes, numerous obstacles come to the forefront [1]. Among these, interconnects emerge as a formidable challenge which demands attention to sustain the scaling endeavor [2]. Various promising designs and materials for interconnects present potential solutions [3], [4], aiming to enhance interconnect performance. Nonetheless, interconnect scaling remains a pivotal concern, validated by many studies predicting the interconnects as performance bottlenecks in advanced technology nodes due to multiple scaling challenges [1], [2]. To find future interconnect candidates, a deep understanding and modeling of interconnect conductivity and scattering mechanisms assume high importance.

Surface scattering is an important mechanism that governs the increase in resistivity of interconnects with the miniaturization of the interconnect cross-sectional area. The well-known Fuchs-Sondheimer (FS) theory [5] models surface scattering using fundamental physical equations. An alternative method based on kinetic-theory principles was suggested in [6] to treat surface scattering. While these methods yield precise results for wires under fully diffusive scattering (specularity $p$=0), obtaining an exact solution for rectangular wires with a general $p$ value has proved challenging [7]. To address this, the approach presented in [6] suggested an infinite series expansion using exact solutions for fully diffusive scattering. A recent empirical approach using "cosh" function [8] provides a spatial-dependent resistivity model. This approach can be useful in modeling modern interconnect vias with a taper structure [9] (i.e. wider openings at the top and narrower openings at the bottom of the interconnects). However, the model, being empirical, lacks physical insights and relies on fitting parameters. In order to achieve both the spatial dependency and explicit relations to the physical parameters, we proposed a spatially resolved FS (SRFS) model derived from the Boltzmann transport equation (BTE) in Part I of this work [10]. However, like the FS model, SRFS model involves computation of complicated integrals, which makes it unsuitable for integrating in computer-aided-design (CAD) software or circuit simulators. Thus, a circuit-compatible model for conductivity accounting for surface scattering is needed, which offers spatial resolution and direction relationships to the interconnect material/geometry parameters.

In this part of our paper, we address this need by building upon the SRFS model of Part I and proposing circuit-compatible conductivity model for rectangular interconnects. The proposed approach models the conductivity (accounting for surface scattering) as a function of the location within the cross-section as well as physical parameters such as electron mean free path ($\lambda_0$), specularity ($p$) and interconnect width/height.

The contributions of Part II of this work are as follows:

- We present a circuit-compatible model for spatially resolved conductivity of rectangular interconnects based on the SRFS model developed in part I [10]. We refer to our model as "SRFS-based Circuit-Compatible Conductivity"

This work was supported , in part, by the NEW materials for LogIc, Memory and InTerconnectS (NEWLIMITS) Center funded by the Semiconductor Research Corporation (SRC)/National Institute of Standards and Technology (NIST) under Award 70NANB17H041. (Corresponding author: Xinkang Chen.)

Xinkang Chen and Sumeet Kumar Gupta are with the School of Electrical and Computer Engineering, Purdue University, West Lafayette, IN 47907 USA (e-mail: chen3030@purdue.edu; guptask@purdue.edu).

$$\frac{\sigma_{SRFS}(x_n, y_n)}{\sigma_0} = \frac{3}{4\pi}\int_{\theta=0}^{\pi} \eta(x_n, y_n, \theta) * cos^2\theta sin\theta d\theta$$

$$\eta(x_n, y_n, \theta) = 2\pi - (1-p) \times \left[ 2\int_{\Phi=0}^{\frac{\pi}{2}} \left( \frac{\left( e^{\frac{-\kappa_a}{2sin\theta cos\Phi}} \times \cosh\left(\frac{\kappa_a \times x_n}{2sin\theta cos\Phi}\right)\right)}{1 - pe^{\frac{-\kappa_a}{sin\theta cos\Phi}}} + \frac{\left( e^{\frac{-\kappa_b}{2sin\theta sin\Phi}} \times \cosh\left(\frac{\kappa_b \times y_n}{2sin\theta sin\Phi}\right)\right)}{1 - pe^{\frac{-\kappa_b}{sin\theta sin\Phi}}} \right) d\Phi + \frac{1}{2}\sum_{n,d}\int_{\Phi=0}^{\frac{\pi}{2}} \left| \frac{e^{\frac{-\kappa_a \times d}{2sin\theta cos\Phi}}}{1 - pe^{\frac{-\kappa_a}{sin\theta cos\Phi}}} - \frac{e^{\frac{-\kappa_b \times n}{2sin\theta sin\Phi}}}{1 - pe^{\frac{-\kappa_b}{sin\theta sin\Phi}}} \right| d\Phi \right]$$

$$\text{Where } (n,d) \rightarrow \begin{Bmatrix} (1+y_n, 1+x_n), (1-y_n, 1+x_n), \\ (1+y_n, 1-x_n), (1-y_n, 1-x_n) \end{Bmatrix} \qquad x_n = \frac{x}{a/2} \in [-1,1], \; y_n = \frac{y}{b/2} \in [-1,1]; \qquad \kappa_a = \frac{a}{\lambda_0}, \kappa_b = \frac{b}{\lambda_0} \tag{1}$$

model or SRFS-C3 model.

- We propose well-behaved functions for the parameters of the SRFS-C3 model relating them to $\lambda_0$ and $p$.
- We extensively validate our circuit-compatible model with the SRFS model showing an excellent match with respect to the spatial profile as well as trends with varying interconnect width/height, $\lambda_0$ and $p$.
- We compare the SRFS-C3 model with the previously proposed "cosh" model [8] (appropriately modified for fair comparison) highlighting the advantages of our approach.
- We present a continuous semi-analytical formula for the average conductivity and validate it against previous FS-based models.

## II. Background

### A. Spatially Resolved FS (SRFS) Model

In Part I of this work [10], we proposed a spatially resolved FS (SRFS) model ($\sigma_{SRFS}$ in equation (1)) derived from the basic Boltzmann transport equation. The main advantage of the SRFS model is that conductivity is expressed as a function of the location within the interconnect cross-section. Further, as in the original FS model, the SRFS model retains the explicit relationships of conductivity to fundamental electron transport parameters viz. $\lambda_0$ and $p$ and the cross-section width ($a$) and height ($b$). As can be observed from (1), the conductivity is a function of $x_n$ and $y_n$ (the x- and y- locations normalized to $a$ and $b$, respectively), the bulk copper conductivity ($\sigma_0$), specularity ($p$, where $0<p<1$ and $p = 0$ for diffusive and $p = 1$ for completely elastic scattering) and the parameters $\kappa_a(= a/\lambda_0)$ and $\kappa_b(= b/\lambda_0)$.

The SRFS model contains nested integrals, and thus its evaluation requires considerable computing resources. To use a conductivity model suitable for CAD software and circuit simulators, the computing requirements of the SRFS model need to be reduced, while retaining its spatial resolution and explicit relations to the physical parameters. The proposed SRFS-C3 model aims to meet these requirements.

### B. The "cosh" Model with Spatial Dependency

A semi-empirical resistivity model (2) proposed in [8] can capture the impact of surface scattering, based on the conductor's dimensions and location within the cross-section.

$$\rho(x,y) = \rho_0 + \rho_q\left(\frac{cosh[\frac{x}{\lambda_q}]}{cosh[\frac{a}{2\lambda_q}]} + \frac{cosh[\frac{y}{\lambda_q}]}{cosh[\frac{b}{2\lambda_q}]}\right) \tag{2}$$

Here, $\rho_0$ lumps bulk resistivity and other scattering mechanisms (such as grain boundary scattering) and the rest of the expression accounts for spatial dependence of resistivity due to surface scattering, with $\rho_q$ and $\lambda_q$ as the model fitting parameters. The main advantage of this approach (referred to as the "cosh" model subsequently) compared to the traditional FS model is its spatial resolution of resistivity. However, this method does not directly capture the relationship of the conductivity with the physical parameters. We have also shown in Part I [10] that the spatial profile predicted by this empirical approach (based on the parameters in [8]) is quite different compared to the physical SRFS model (derived from fundamental equations), even though the average conductivity value for the two models is the same (and matched to the experiments in [8]). However, the question is if the "cosh" function can be used in conjunction with the SRFS model for better prediction of the spatial profile of the conductivity. We explore this later in the paper.

## III. Spatially Resolved FS-based Circuit-compatible Conductivity (SRFS-C3) Model

As noted before, the SRFS model proposed in Part I [10] provides spatial resolution as well the connections to the physical parameters but lacks circuit compatibility. In other words, if the conductivity at every point in the cross-section of the interconnect is calculated using the SRFS model, the time of computation would be prohibitively large for a circuit or CAD simulation. To get around this problem, the basic premise of our SRFS-C3 model is to develop an analytical function which can predict the spatial profile of the conductivity based on the conductivity values obtained from the SRFS model for only a *few* spatial locations. As we will show later, the SRFS-C3 approach requires the SRFS conductivity values ($\sigma_{SRFS}$) from (1) for only 4 points for a rectangular interconnect (and only 3 points for a symmetric square interconnect). Thus, by calling the SRFS model only four times and using the proposed analytical function, the spatial profile for the entire cross-section as well as the trends with respect to varying $p$, $\lambda_0$, $a$ and $b$ can be predicted by SRFS-C3 with a good accuracy. This helps in significantly speeding the computation, thus making the SRFS-C3 model circuit compatible. To explain our technique, let us first start with the 1D SRFS-C3 model for the spatially-resolved conductivity. We will then extend it to 2D.

### A. 1D SRFS-C3 Model

While analyzing the SRFS model using (1), we observe that the partial derivative of the spatially-resolved conductivity ($\sigma$)

with respect to $x$ (or $y$) has a profile similar to the $tanh^{-1}$ function (especially near the center of the interconnect). However, if we integrate the $tanh^{-1}$ function to find the conductivity, the resultant model is unable to predict the sharp slopes closer to the edges of the interconnect, especially for high values of $\kappa_a$. On the other hand, the cosh function used in [8] tracks the sharpness of the spatial profile near the edges for a wide range of $\kappa_a$, but is unable to match the overall space dependence (as also shown in Part I [10]). Therefore, in order to get a good match in the entire the spatial region, we proceed by combining the features of $tanh^{-1}$ and *cosh* functions as follows. We start with the second partial derivative of $\sigma$ with respect to $x_n$ to reflect the functional attributes of $tanh^{-1}$ based conductivity derivative combined with *cosh*-based conductivity profile. For that, we use the following analytical function.

$$\frac{\partial^2 \sigma}{\partial x_n^2} = \frac{\alpha_x \cosh(\beta_a x_n)}{1 - {x_n}^2} \tag{3}$$

Here, *cosh*-function is used to represent the double derivative of *cosh*-based conductivity profile and $\frac{1}{1-{x_n}^2}$ is used to represent the derivative of *tanh*-based conductivity derivative. The two are multiplied to combine the attributes of the respective functions. The parameters $\alpha_x$ and $\beta_a$ will be discussed in detail subsequently.

To calculate the conductivity, we perform a double integral of (3) with respect to $x_n$, use the boundary conditions at $x_n$=0 (center of the wire) and $x_n$=1 (edge of the wire) and obtain

$$\sigma(x_n) = \sigma|_{x_n=0} - \left(\sigma|_{x_n=0} - \sigma|_{x_n=1}\right) \times \frac{\xi(x_n, \beta_a)}{\xi_1(\beta_a)} \tag{4}$$

Where $\xi(x_n, \beta_a) =$

$$(1 + x_n)\left\{\begin{matrix} cosh(\beta_a)\left[Chi\left(\beta_a(1 + x_n)\right) - Chi(\beta_a)\right] \\ -sinh(\beta_a)\left[Shi\left(\beta_a(1 + x_n)\right) - Shi(\beta_a)\right] \end{matrix}\right\} + (1 - x_n)\left\{\begin{matrix} cosh(\beta_a)\left[Chi\left(\beta_a(1 - x_n)\right) - Chi(\beta_a)\right] \\ -sinh(\beta_a)\left[Shi\left(\beta_a(1 - x_n)\right) - Shi(\beta_a)\right] \end{matrix}\right\} \tag{5}$$

And $\xi_1(\beta_a) = \xi(1, \beta_a)$

$$= 2 \times \left[cosh(\beta_a)\left(Chi(2\beta_a) - Chi(\beta_a)\right) - sinh(\beta_a)\left(Shi(2\beta_a) - Shi(\beta_a)\right)\right] \tag{6}$$

Here, *Chi* is the cosh-integral (‘coshint’ in MATLAB) and *Shi* is the sinh-integral ('sinhint' in MATLAB) functions. The parameter $\beta_a$ is used for fitting and is obtained by minimizing the mean square error (MSE) between the circuit-compatible and the SRFS model. $\beta_a$ is a function of $\kappa_a$ (and therefore, $a$ and $\lambda_0$) and $p$. We will discuss the $\beta_a$ fitting and an analytical model for $\beta_a$ in detail later. The parameter $\alpha_x$ is assimilated in the expressions above once the boundary conditions are applied. The boundary values $\sigma|_{x_n=0}$ and $\sigma|_{x_n=1}$ need to be obtained from the 1D SRFS model i.e. using (1) with $y_n$-dependent terms ignored (e.g. by setting $\kappa_b \rightarrow \infty$, see Part I [10] for details). Note, the spatial profile is expected to be symmetric with respect to the origin, $\sigma|_{x_n=1} = \sigma|_{x_n=-1}$, and therefore, we use the boundary conditions at the center and edge of the wire along the $x$ direction. Thus, by using just 2 values from the 1D SRFS model, the proposed 1D SRFS-C3 model in (4) can predict the spatial profile for any $x_n$ (in the range of -1 and 1). We can repeat the same procedure for obtaining the spatial dependence with respect to $y_n$ with its own fitting parameter $\beta_b$ (function of $p$ and $\kappa_b$ (and hence, $b$ and $\lambda_0$)) and boundary points $\sigma|_{y_n=0}$ and $\sigma|_{y_n=1}$ obtained from the SRFS model. With this discussion in mind, let us now explain the 2D SRFS-C3 model.

### B. 2D SRFS-C3 Model

To obtain the 2D SRFS-C3 model from the 1D model of (4), we use the following procedure:

- Given an arbitrary location ($x_n$, $y_n$) within the interconnect cross-section at which we want to find $\sigma(x_n, y_n)$, we first use (4) to obtain

$$\sigma(x_n, 0) = \sigma_{SRFS}(0,0) - \left(\sigma_{SRFS}(0,0) - \sigma_{SRFS}(1,0)\right) \times \frac{\xi(x_n, \beta_a)}{\xi_1(\beta_a)} \tag{7}$$

Here, $\sigma_{SRFS}$ refers to the SRFS model derived in Part I [10] and given by (1).

- Following a similar process, we obtain $\sigma(x_n, 1)$ as

$$\sigma(x_n, 1) = \sigma_{SRFS}(0,1) - \left(\sigma_{SRFS}(0,1) - \sigma_{SRFS}(1,1)\right) \times \frac{\xi(x_n, \beta_a)}{\xi_1(\beta_a)} \tag{8}$$

- Then, we use the 1D equation along $y$ to calculate $\sigma(x_n, y_n)$

$$\sigma(x_n, y_n) = \sigma(x_n, 0) - \left(\sigma(x_n, 0) - \sigma(x_n, 1)\right) \times \frac{\xi(y_n, \beta_b)}{\xi_1(\beta_b)} \tag{9}$$

Here, $\sigma(x_n, 0)$ and $\sigma(x_n, 1)$ are obtained from (7) and (8).

We follow the process described above to obtain an analytical formula for $\sigma(x_n, y_n)$

$$\begin{aligned} \sigma(x_n, y_n) = {} & \sigma_{SRFS}(0,0) \\ & -\left(\sigma_{SRFS}(0,0) - \sigma_{SRFS}(1,0)\right) \times \frac{\xi(x_n, \beta_a)}{\xi_1(\beta_a)} \\ & -\left(\sigma_{SRFS}(0,0) - \sigma_{SRFS}(0,1)\right) \times \frac{\xi(y_n, \beta_b)}{\xi_1(\beta_b)} \\ & -\left(\sigma_{SRFS}(0,1) + \sigma_{SRFS}(1,0) - \sigma_{SRFS}(0,0) - \sigma_{SRFS}(1,1)\right) \\ & \times \frac{\xi(x_n, \beta_a)}{\xi_1(\beta_a)} \frac{\xi(y_n, \beta_b)}{\xi_1(\beta_b)} \end{aligned} \tag{10}$$

It can be noted that we need only four points from the SRFS model of (1) to compute the entire spatial profile. This is made possible by the analytical $\xi$ function described by (5), which forms the basis of the proposed SRFS-C3 model. It is also noteworthy that (10) comprises of terms with sole dependences on $x_n$ (2[nd] term) or $y_n$ (3[rd] term), but additionally it also captures the cross interactions through the last term which has the product of $x_n$- and $y_n$-dependent terms. This is similar to the SRFS model in equation (1) which has sole $x_n$ and $y_n$-dependent terms and a term which depends on both $x_n$ and $y_n$.

### C. SRFS-C3 Parameter $\beta_a(\beta_b)$: Fitting and Modeling

The parameter $\beta_a(\beta_b)$ can be obtained by minimizing the mean square error (MSE) between the SRFS-C3 and SRFS models. To keep the model simple, we perform the fitting independently for $\beta_a$ and $\beta_b$, that is by neglecting the cross interaction between $x_n$ and $y_n$ for the fitting parameters. (Note, however, that the $x_n$-$y_n$ interaction in the conductivity profile *is* captured by the last term in (10), even though $\beta_a$ and $\beta_b$ are obtained independently). We will show later that despite this assumption for the fitting parameters, the SRFS-C3 models shows a close match to the SRFS model. As discussed before, $\beta_a(\beta_b)$ is a function of $\kappa_a(\kappa_b)$ and $p$. Hence, we find the best fit for $\beta_a$ by sweeping $\kappa_a$ and $p$. (Note, by relating $\beta_a$ to $\kappa_a$, we can seamlessly find $\beta_b$ by substituting $\kappa_b$ for $\kappa_a$. Henceforth, we will present our discussions in terms of $\beta_a$ and $\kappa_a$).The best fit values for $\beta_a$ for different $\kappa_a$ and $p$ shown in Fig. 1. Interestingly, for $\kappa_a$>1, $\beta_a$ is independent of $p$ and exhibits a linear behavior with respect to $\kappa_a$ with a slope of $\frac{2}{\pi}$. On the other hand, for small $\kappa_a$ ($\kappa_a$<1), $\beta_a$ shows a deviation from this linear behavior and exhibits a dependence on $p$. Further, at $p$=0 and $\kappa_a \to 0$, $\beta_a \to 1$. Based on these empirical observations, the relationship of $\beta_a$ with $\kappa_a$ and $p$ can be approximately captured using the following equation:

$$\beta_a = \max\left[\frac{2}{\pi}\kappa_a + 1 + \ln\left(\tanh\frac{\kappa_a^2}{\varepsilon p}\right), 0.01\right] \tag{11}$$

Here, $\varepsilon$ is a fitting parameter and is found to be equal to 0.42 by minimizing the MSE between (11) and the best fit values of $\beta_a$. Fig. 1a-d illustrates the behavior of (11) for different $p$ and $\kappa$ values, demonstrating that (11) can model the relationship between $\beta_a$, $\kappa_a$ and $p$ with reasonable accuracy using a simple equation (except for some deviations for small $\kappa_a$).

The complete SRFS-C3 model is summarized at the bottom of this page. Note, the SRFS-C3 model is based on conductivity subtraction which is more physical as we noted in Part I [10] compared to resistivity addition used in the "cosh" model [8].

The SRFS-C3 model is general and can also be applied to thin films (of thickness $b$ along the y-direction). For that, we set $\kappa_a \to \infty$ and obtain

$$\sigma_{SRFS-C3}(y_n) = \sigma_{SRFS}(0,0) - \left(\sigma_{SRFS}(0,0) - \sigma_{SRFS}(0,1)\right) \times \frac{\xi(y_n, \beta_b)}{\xi_1(\beta_b)} \tag{12}$$

Note that one can also think of using polynomial and 2D Fourier interpolation methods, by using some points from the SRFS calculation and predicting the spatial profile based on them. However, that will require the calculation of the SRFS multiples times, which reduces the circuit compatibility. Implementing polynomial or Fourier interpolation with just four SRFS points does not adequately capture the spatial profile. Furthermore, since the shape of the conductivity profile is highly dependent on $\kappa$ (and to some extent $p$), the parameters of the polynomial interpolation (for example optimal number of points, or the sampling points) may depend on $\kappa$, which would reduce the adaptability of the model and its seamless application for a wide range of parameters.

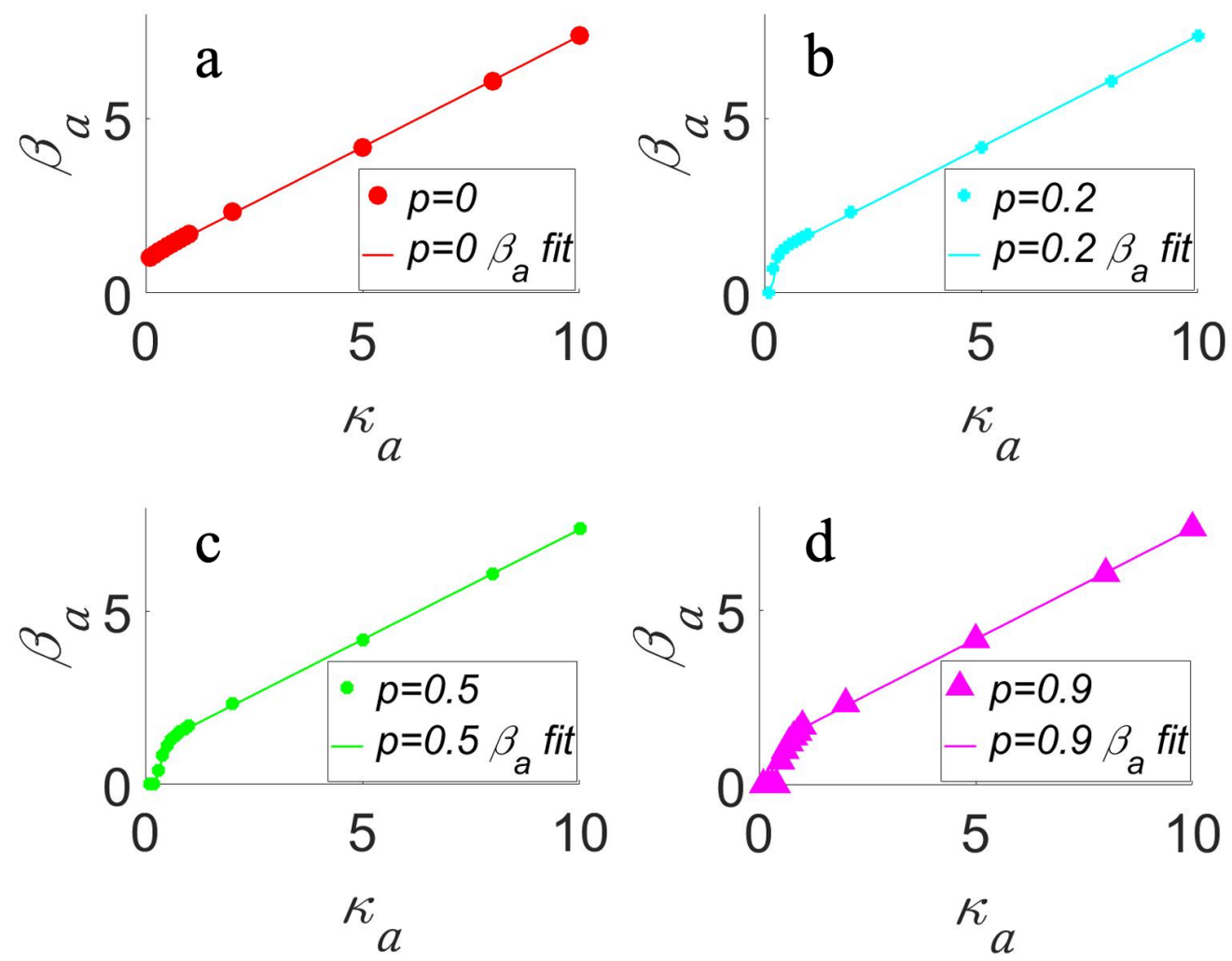

Fig. 1. $\beta_a$ versus $\kappa_a$ showing the best-fit values in dots and approximate-fit values obtain from the model in (11) in lines for a) *p=0* b) *p=0.2* c) *p=0.5* d) *p=0.9*

However there is one question to explore: Can the "cosh" model (similar to [8]) be used in conjunction with only a few values obtained from the SRFS model (similar to our approach) to predict spatial profile of the conductivity. We explore this question next.

## IV. Modified "cosh" Model

In Part I of this work [10], we compared our proposed SRFS conductivity model and the "cosh" model from [8] by matching their average conductivity. We observe significant differences in their spatial profiles. Therefore, for a fair comparison of the "cosh" model with the SRFS-C3 approach, in this section, we use the "cosh" model in conjunction with the SRFS model, following a similar approach as the SRFS-C3 model.

We utilize the "cosh" model as the analytical function to predict the spatial profile of restivity based on a few resistivity (1/conductivity) values obtained from the SRFS model i.e. the reciprocal of equation (1). Following the similar approach as the last sub-section, we find four values from the SRFS model - $\sigma_{SRFS}(0,0)$, $\sigma_{SRFS}(1,0)$, $\sigma_{SRFS}(0,1)$ and $\sigma_{SRFS}(1,1)$. The modified normalized "cosh" model is presented in (13).

Final SRFS-C3 model for conductivity accounting for specular surface scattering

$$\sigma_{SRFS-C3}(x_n, y_n) = \sigma_{SRFS}(0,0) - \left(\sigma_{SRFS}(0,0) - \sigma_{SRFS}(1,0)\right) \times \frac{\xi(x_n, \beta_a)}{\xi_1(\beta_a)} - \left(\sigma_{SRFS}(0,0) - \sigma_{SRFS}(0,1)\right) \times \frac{\xi(y_n, \beta_b)}{\xi_1(\beta_b)}$$
$$-\left(\sigma_{SRFS}(0,1) + \sigma_{SRFS}(1,0) - \sigma_{SRFS}(0,0) - \sigma_{SRFS}(1,1)\right) \times \frac{\xi(x_n, \beta_a)}{\xi_1(\beta_a)} \frac{\xi(y_n, \beta_b)}{\xi_1(\beta_b)}$$

where

$$\xi(x_n, \beta_a) = (1 + x_n)\left\{cosh(\beta_a)\left[Chi\left(\beta_a(1 + x_n)\right) - Chi(\beta_a)\right] - sinh(\beta_a)\left[Shi\left(\beta_a(1 + x_n)\right) - Shi(\beta_a)\right]\right\}$$
$$+(1 - x_n)\left\{cosh(\beta_a)\left[Chi\left(\beta_a(1 - x_n)\right) - Chi(\beta_a)\right] - sinh(\beta_a)\left[Shi\left(\beta_a(1 - x_n)\right) - Shi(\beta_a)\right]\right\}$$
$$\xi(y_n, \beta_b) = (1 + y_n)\left\{cosh(\beta_b)\left[Chi\left(\beta_b(1 + y_n)\right) - Chi(\beta_b)\right] - sinh(\beta_b)\left[Shi\left(\beta_b(1 + y_n)\right) - Shi(\beta_b)\right]\right\}$$
$$+(1 - y_n)\left\{cosh(\beta_b)\left[Chi\left(\beta_b(1 - y_n)\right) - Chi(\beta_b)\right] - sinh(\beta_b)\left[Shi\left(\beta_b(1 - y_n)\right) - Shi(\beta_b)\right]\right\}$$

and

$$\xi_1(\beta_a) = 2 \times \left[cosh(\beta_a)\left(Chi(2\beta_a) - Chi(\beta_a)\right) - sinh(\beta_a)\left(Shi(2\beta_a) - Shi(\beta_a)\right)\right]$$
$$\xi_1(\beta_b) = 2 \times \left[cosh(\beta_b)\left(Chi(2\beta_b) - Chi(\beta_b)\right) - sinh(\beta_b)\left(Shi(2\beta_b) - Shi(\beta_b)\right)\right]$$
$$x_n = \frac{x}{a/2} \in [-1,1],\ \ y_n = \frac{y}{b/2} \in [-1,1];\ \ \kappa_a = \frac{a}{\lambda_0},\ \kappa_b = \frac{b}{\lambda_0};$$
$$\beta_a = max\left[\frac{2}{\pi}\kappa_a + 1 + \ln\left(\tanh\frac{\kappa_a^2}{0.42p}\right), 0.01\right], \qquad \beta_b = max\left[\frac{2}{\pi}\kappa_b + 1 + \ln\left(\tanh\frac{\kappa_b^2}{0.42p}\right), 0.01\right]$$

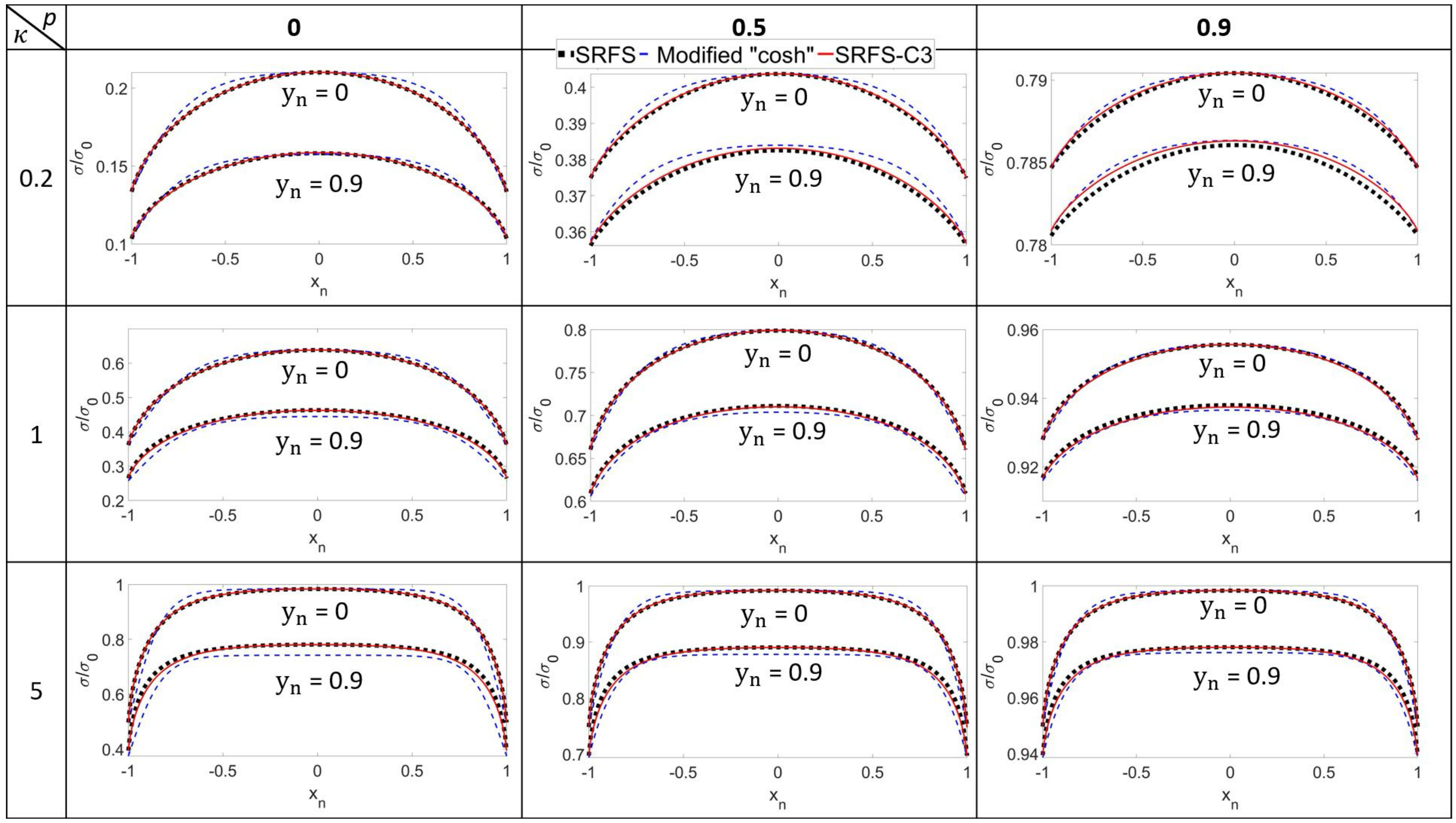

Fig. 2. The spatial profiles for $\sigma/\sigma_0$ (conductivity normalized to bulk conductivity) versus $x_n$ ($=x/(a/2)$) for $y_n$ ($=y/(b/2)$) = 0, and 0.9 for different $\kappa$ and $p$. The model parameters correspond to the approximate fitting scenario.

$$\rho(x_n, y_n) = \rho_b + \rho_{qx}\left(\frac{\cosh\left[\frac{\kappa_a x_n}{2\lambda_{na}}\right]-1}{\cosh\left[\frac{\kappa_a}{2\lambda_{na}}\right]-1}\right) + \rho_{qy}\left(\frac{\cosh\left[\frac{\kappa_b y_n}{2\lambda_{nb}}\right]-1}{\cosh\left[\frac{\kappa_b}{2\lambda_{nb}}\right]-1}\right)$$
$$-\rho_q\left(\frac{\cosh\left[\frac{\kappa_a x_n}{2\lambda_{na}}\right]-1}{\cosh\left[\frac{\kappa_a}{2\lambda_{na}}\right]-1}\right)\times\left(\frac{\cosh\left[\frac{\kappa_b y_n}{2\lambda_{nb}}\right]-1}{\cosh\left[\frac{\kappa_b}{2\lambda_{nb}}\right]-1}\right) \quad (13)$$

$$\rho_b = \frac{1}{\sigma_{SRFS}(0,0)}, \rho_{qx} = \frac{1}{\sigma_{SRFS}(1,0)} - \rho_b, \rho_{qy} = \frac{1}{\sigma_{SRFS}(0,1)} - \rho_b$$
$$\rho_q = \frac{1}{\sigma_{SRFS}(1,0)} + \frac{1}{\sigma_{SRFS}(0,1)} - \frac{1}{\sigma_{SRFS}(0,0)} - \frac{1}{\sigma_{SRFS}(1,1)}$$

Here $\lambda_{na}(=\lambda_{qa}/\lambda_0)$ and $\lambda_{nb}(=\lambda_{qb}/\lambda_0)$ are the new model fitting parameters (recall, $\lambda_q$ is the fitting parameter in (2) from [8]). We find the best fitting value for $\lambda_{na}(\lambda_{nb})$ based on minimizing the MSE with respect to the SRFS model. We propose an approximate fitting expression as follows:

$$\lambda_{na} = 0.234 * tanh(0.4\kappa_a)$$
$$+0.0875 * (tanh(0.7\kappa_a) + 0.03) * p \quad (14)$$

The $\lambda_{na}$ expression in (14) is a bit more complex compared to $\beta_a$ of (11) used in the proposed SRFS-C3 model. Moreover, the accuracy of the fit for (14) is also not as good as (11). One potential option to improve the accuracy is to consider $\lambda_{na}$ (and $\lambda_{nb}$) as function of both $\kappa_a$ and $\kappa_b$, albeit at the cost of increasing the model complexity. Our analysis shows that this improves the model accuracy but only very slightly. Hence, to keep the model simple, we model $\lambda_{na}$ solely as a function of $\kappa_a$ ($\lambda_{nb}$ solely as a function of $\kappa_b$), as per (14). Also, as we have pointed out before in section III, the modified "cosh" model is based on resistivity addition. In contrast, the proposed SRFS-C3 is based on conductivity subtraction, which is more consistent with the physical models of the original FS [5] and our SRFS approach.

## V. Analysis of the Spatial Profile of Conductivity

In this section, we compare the spatial resolution of conductivity of a square interconnect ($\kappa_a$=$\kappa_b$=$\kappa$) predicted by the proposed SRFS-C3 model, the modified "cosh" model and the accurate physical SRFS model for three $\kappa$ values (0.2, 1 and 5) and three $p$ values (0, 0.5 and 0.9), covering a wide range of these parameters. (We have made similar comparisons for rectangular wires with similar conclusions; here, we do not include the results to avoid repetition).

Fig. 2 shows the comparison with approximate models for the fitting parameters obtained from (11) and (14) incorporated (referred as approximate fit). We show the profiles as function of $x_n$ for two different values of $y_n$. Note that since the SRFS-C3 and SRFS models are symmetric with respect to $x_n$ and $y_n$, we observe similar profiles when we plot the conductivity profiles with respect to $y_n$ (not shown here in the interest of space). It can be observed that the SRFS-C3 model offers an excellent match to the physical SRFS model across a wide range of $\kappa_a$ and $p$. Furthermore, compared to the modified "cosh" model, the proposed SRFS-C3 exhibits a better match with the SRFS model. The only exception is at very small $\kappa_a$ and high $p$ where the modified "cosh" performs slightly better.

For $y_n$ = 0.9, the limitations of the modified "cosh" model become even more evident. On the other hand, the SRFS-C3 model shows a close match to the SRFS for both $y_n$ = 0 and 0.9.

To show how well the SRFS-C3 and modified "cosh" models emulate the conductivity profile obtain from SRFS, we report the error between the SRFS-C3 and SRFS, and modified "cosh" and SRFS models considering the entire 2D cross-section of the interconnect. We show the results for two cases: (a) best fit (smallest MSE for $\beta_a$ and $\lambda_{na}$ without using (11) and (14)) in

TABLE I

Average error across the 2D cross-section between SRFS-C3 and SRFS, and modified “cosh” and SRFS model (best fit)

| P \ $\kappa$ | 0.2 | | 0.5 | | 1 | | 2 | | 5 | | 10 | |
|---|---|---|---|---|---|---|---|---|---|---|---|---|
| | SRFS-C3 | Modified “cosh” | SRFS-C3 | Modified “cosh” | SRFS-C3 | Modified “cosh” | SRFS-C3 | Modified “cosh” | SRFS-C3 | Modified “cosh” | SRFS-C3 | Modified “cosh” |
| **0** | 0.2% | 1.4% | 0.2% | 1.8% | 0.3% | 2.4% | 0.4% | 3.2% | 0.3% | 3.5% | 0.3% | 2.4% |
| **0.5** | 0.1% | 0.1% | 0.0% | 0.3% | 0.0% | 0.5% | 0.1% | 0.9% | 0.1% | 1.2% | 0.1% | 0.9% |
| **0.9** | 0.1% | 0.0% | 0.0% | 0.1% | 0.0% | 0.2% | 0.1% | 0.5% | 0.3% | 0.8% | 0.3% | 0.8% |

TABLE II

Average error across the 2D cross-section between SRFS-C3 and SRFS, and modified “cosh” and SRFS model (approximate fit)

| P \ $\kappa$ | 0.2 | | 0.5 | | 1 | | 2 | | 5 | | 10 | |
|---|---|---|---|---|---|---|---|---|---|---|---|---|
| | SRFS-C3 | Modified “cosh” | SRFS-C3 | Modified “cosh” | SRFS-C3 | Modified “cosh” | SRFS-C3 | Modified “cosh” | SRFS-C3 | Modified “cosh” | SRFS-C3 | Modified “cosh” |
| **0** | 0.1% | 2.8% | 0.3% | 2.3% | 0.5% | 2.4% | 0.7% | 3.2% | 0.3% | 3.5% | 0.3% | 2.4% |
| **0.5** | 0.1% | 0.8% | 0.1% | 0.6% | 0.2% | 0.6% | 0.2% | 0.9% | 0.1% | 1.2% | 0.1% | 0.9% |
| **0.9** | 0.0% | 0.1% | 0.0% | 0.1% | 0.1% | 0.1% | 0.1% | 0.1% | 0.0% | 0.2% | 0.0% | 0.2% |

TABLE III

Maximum error across the 2D cross-section between SRFS-C3 and SRFS, and modified “cosh” and SRFS model (best fit)

| P \ $\kappa$ | 0.2 | | 0.5 | | 1 | | 2 | | 5 | | 10 | |
|---|---|---|---|---|---|---|---|---|---|---|---|---|
| | SRFS-C3 | Modified “cosh” | SRFS-C3 | Modified “cosh” | SRFS-C3 | Modified “cosh” | SRFS-C3 | Modified “cosh” | SRFS-C3 | Modified “cosh” | SRFS-C3 | Modified “cosh” |
| **0** | 0.9% | 5.5% | 1.2% | 7.7% | 1.8% | 10.0% | 3.0% | 15.9% | 5.1% | 22.0% | 6.0% | 23.1% |
| **0.5** | 0.3% | 0.3% | 0.1% | 1.0% | 0.2% | 1.8% | 0.8% | 3.3% | 1.6% | 5.4% | 2.0% | 5.9% |
| **0.9** | 0.1% | 0.0% | 0.0% | 0.1% | 0.0% | 0.2% | 0.1% | 0.5% | 0.3% | 0.8% | 0.3% | 0.8% |

TABLE IV

Maximum error across the 2D cross-section between SRFS-C3 and SRFS, and modified “cosh” and SRFS model (approximate fit)

| P \ $\kappa$ | 0.2 | | 0.5 | | 1 | | 2 | | 5 | | 10 | |
|---|---|---|---|---|---|---|---|---|---|---|---|---|
| | SRFS-C3 | Modified “cosh” | SRFS-C3 | Modified “cosh” | SRFS-C3 | Modified “cosh” | SRFS-C3 | Modified “cosh” | SRFS-C3 | Modified “cosh” | SRFS-C3 | Modified “cosh” |
| **0** | 0.9% | 6.4% | 1.4% | 5.8% | 2.3% | 10.0% | 3.6% | 16.7% | 5.0% | 22.6% | 6.1% | 23.1% |
| **0.5** | 0.3% | 1.5% | 0.2% | 1.3% | 0.5% | 1.8% | 1.0% | 3.7% | 1.6% | 5.4% | 2.0% | 5.9% |
| **0.9** | 0.1% | 0.1% | 0.0% | 0.2% | 0.1% | 0.2% | 0.2% | 0.6% | 0.3% | 0.8% | 0.3% | 0.8% |

Table I and (b) approximate fit cases (using (11) and (14)) in Table II. Maximum error across the whole interconnect cross section is also compared in Table III and Table IV. For the best fit case, the average error for SRFS-C3 is < 0.4 % across all $\kappa$ and $p$, and the maximum error is < 6%. On the other hand, for modified "cosh", the average error < 3.5%, and the maximum error is < 23.1%.

We also quantify the speed-up obtained with the circuit-compatible models over the SRFS model. For that, we evaluate a surface with a grid of 201 by 201 sampling points across the interconnect cross-section. The baseline is the SRFS model using MATLAB’s in-built integration function and ‘parfor’ loops. As an alternate method, we implement a 10-point Gaussian quadrature scheme for the SRFS model, which is 7.3 times faster than the baseline. On the other hand, the proposed SRFS-C3 model achieves a much larger speed-up of ~48.6X over the baseline. We note that the computation time of the SRFS-C3 model can be further improved by forming look-up tables for the *Chi* and *Shi* functions (instead of utilizing the MATLAB functions to compute them). Note, such a look-up table based approach is feasible since *Chi* and *Shi* are well-defined functions. Also note, the same look-up tables are used independent of the model parameters (such as $\kappa$ and $p$) since we form the look-up table for standard functions. Therefore, this step incurs a negligible one-time cost. Thus, by employing lookup table for the *Chi* and *Shi* functions, rather than recalculating them each time, the speed-up over the baseline is further increased to 81X. The modified "cosh" model also achieves 81X speed-up. The proposed SRFS-C3 model significantly reduces computational time compared to the original SRFS model, while achieving a close match.

## VI. Semi-Analytical Average Conductivity Model derived from SRFS-C3

A remarkable implication of the SRFS-C3 model is that it can yield a continuous analytical model for the average conductivity accounting for surface scattering in rectangular interconnects for a wide range of the physical parameters. As we noted from Part I [10] that in [11], the FS model for square wires ($\kappa_a = \kappa_b$) for $p$=0 was approximated to provide different analytical expressions for different regimes (such as for $\kappa_a \gg 1$, $\kappa_a > 4$, $\kappa_a \ll 1, \kappa_a \sim 1$ etc.). While highly useful in providing compact models for average conductivity and relating them to physical parameters, these expressions do not seamlessly cover the entire range of $\kappa_a$. Some follow-up works have extended these expressions for rectangular wires and for general $p$, but the limitations of these expressions in terms of their validity for only a certain range of $\kappa_a$ still remain. With the SRFS-C3 model, we get the following benefits: (i) A single model is valid for a wide range of $\kappa_a$, $\kappa_b$ and $p$. (ii) The model is analytical and readily integrable and (iii) The model shows a close match with the physical SRFS model. In other words, we can get around the limitations of the approximate analytical expressions from [11] and obtain an accurate continuous model. The only downside compared to the analytical expressions in

$$\sigma_{avg} = \sigma_{SRFS}(0,0) - \left(\sigma_{SRFS}(0,0) - \sigma_{SRFS}(1,0)\right) \times \left[1 - \frac{\frac{\sinh(\beta_a)}{\beta_a}}{\xi_1(\beta_a)}\right] - \left(\sigma_{SRFS}(0,0) - \sigma_{SRFS}(0,1)\right) \times \left[1 - \frac{\frac{\sinh(\beta_b)}{\beta_b}}{\xi_1(\beta_b)}\right]$$

$$- \left(\sigma_{SRFS}(0,1) + \sigma_{SRFS}(1,0) - \sigma_{SRFS}(0,0) - \sigma_{SRFS}(1,1)\right) \times \left[1 - \frac{\sinh(\beta_a)/\beta_a}{\xi_1(\beta_a)}\right] \times \left[1 - \frac{\sinh(\beta_b)/\beta_b}{\xi_1(\beta_b)}\right]$$

where

$$\xi_1(\beta_a) = 2 \times \left[cosh(\beta_a)\left(Chi(2\beta_a) - Chi(\beta_a)\right) - sinh(\beta_a)\left(Shi(2\beta_a) - Shi(\beta_a)\right)\right]$$

$$\xi_1(\beta_b) = 2 \times \left[cosh(\beta_b)\left(Chi(2\beta_b) - Chi(\beta_b)\right) - sinh(\beta_b)\left(Shi(2\beta_b) - Shi(\beta_b)\right)\right]$$

$$\kappa_a = \frac{a}{\lambda_0},\quad \kappa_b = \frac{b}{\lambda_0}, \beta_a = max\left[\frac{2}{\pi}\kappa_a + 1 + \ln\left(\tanh\frac{\kappa_a^2}{0.42p}\right), 0.01\right], \beta_b = max\left[\frac{2}{\pi}\kappa_b + 1 + \ln\left(\tanh\frac{\kappa_b^2}{0.42p}\right), 0.01\right] \quad (15)$$

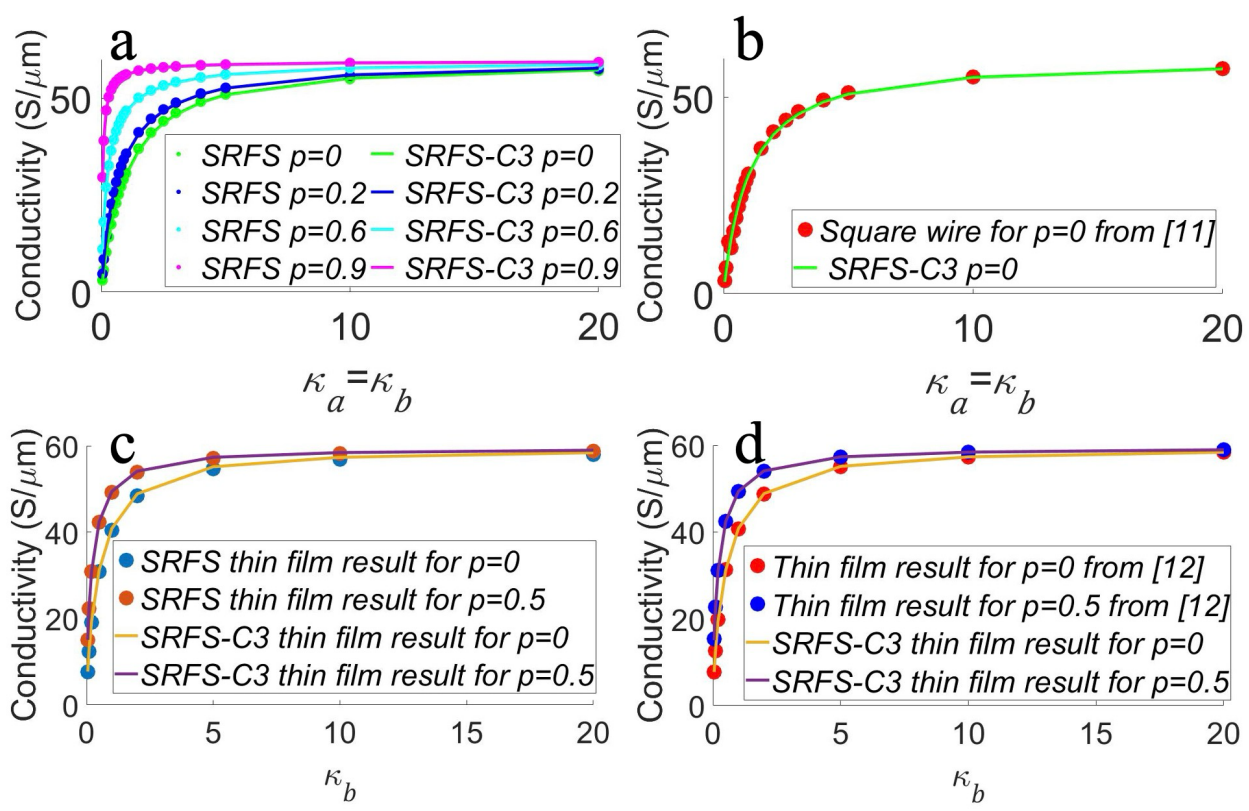

Fig. 3. a) SRFS model versus SRFS-C3 model for square wire with different $\kappa$ and $p$ showing a good match. b) Square wire SRFS-C3 model compared with 2D FS model from [11] for different $\kappa$ showing a close match. c) Thin film SRFS-C3 model versus thin film SRFS for different $\kappa$ and $p$ showing a good match. d) Thin-film SRFS-C3 model compared with thin film FS model from [12] for different $\kappa$ and $p$ showing a close match.

[12] is that the proposed model requires four values by computing the integrals in (1), which makes it semi-analytical.

The model for the average conductivity ($\sigma_{avg}$) obtained by integrating (10) with respect to $x_n$ and $y_n$ (with integral limits of -1 to 1) and dividing by area (= *ab*) is shown in (15).

Fig. 3a compares $\sigma_{avg}$ from (15) with that obtained by numerically integrating the SRFS model and dividing by *ab* for square wires ($\kappa_a$=$\kappa_b$=$\kappa$). The results show a close match across different $\kappa$ and $p$ values. We also compare the average conductivity obtained from (15) with the values from the models in [11] for $p$=0 in Fig. 3b, again showing a close match.

Using the same approach as for the rectangular wire, we obtain the average conductivity ($\sigma_{avg}$) for thin film as

$$\sigma_{avg} = \sigma_{SRFS}(0,0) \\ -\big(\sigma_{SRFS}(0,0) - \sigma_{SRFS}(0,1)\big) \times \left[1 - \frac{\frac{\sinh(\beta_b)}{\beta_b}}{\xi_1(\beta_b)}\right] \qquad (16)$$

(Note for thin films, we set $\kappa_a \rightarrow \infty$ and study the effect of different $\kappa_b$). Our SRFS-C3 model closely matches the results from the SRFS model (Fig. 3c) and the original FS model [12] (Fig. 3d).

An alternate approach to capture surface scattering is based on kinetic theory [7]. The said model uses an infinite series expansion based on average conductivity for $p$=0 and predicts the average conductivity for a general $p$ as shown in (17).

$$\left(\frac{\sigma}{\sigma_0}\right)_{p,\lambda_0} = (1-p)^2 \sum_{j=1}^{\infty} \left\{ j p^{j-1} \left(\frac{\sigma}{\sigma_0}\right)_{p=0,\lambda_0/j} \right\} \qquad (17)$$

Here the $\frac{\sigma}{\sigma_0}$ ratio for general $p$ is define as the infinite weighted sum of $\frac{\sigma}{\sigma_0}$ for $p$=0 and $\lambda_0 = \lambda_0/j$. We compare the result obtained from this approach to the average conductivity obtained from the proposed SRFS-C3 (Fig. 4). As we had noted in Part I [10], the SRFS model uses some approximations for general specularity ($p$), which leads to some differences between the two models. However, the mismatch is less than 10%. Thus, the proposed SRFS-C3 approach is able to model the conductivity for general specularity ($p$) in rectangular interconnects with a reasonable accuracy. While the proposed SRFS-C3 only needs to compute the integrals from SRFS four times to obtain the average conductivity ratio, the model from [7] needs many computations of such integrals (depending on the value of $j$ and the spatial granularity over which the integrals are calculated to obtain the average conductivity for $p$=0). Hence, the proposed SRFS-C3 approach offers a much faster method compared to the infinite weighted sum of (17).

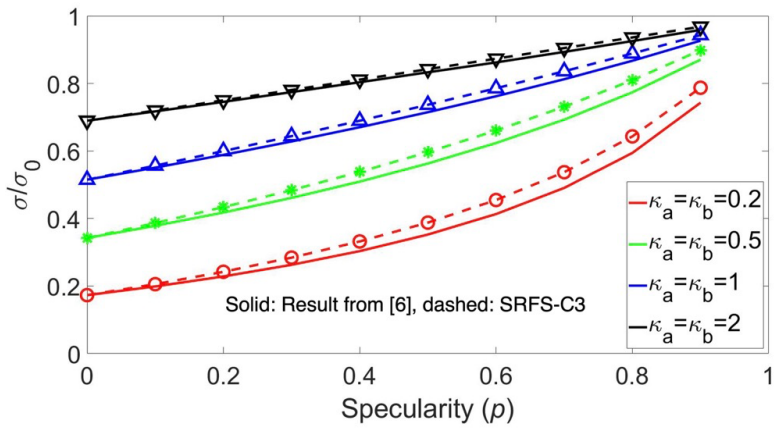

Fig. 4. $\sigma/\sigma_0$ (conductivity normalized to bulk conductivity) for different $p$ and $\kappa$ for a square wire predicted by SRFS-C3 and the model in [6], showing SRFS-C3 model is able to capture conductivity trends for general $p$.

The proposed model in (15) offers a continuous analytical model for average conductivity accounting for surface scattering and providing relationships to physical parameters ($p$, $\lambda_0$, $a$ and $b$). It only requires four computations of integrals given by (1), but despite that, the computation is fast enough for incorporation in circuit simulators and compatible CAD software.

It is noteworthy that if we try computing the average conductivity using the modified "cosh" model, it leads to complicated integrals, for which analytical expressions are challenging to obtain. This is because it is based on resistivity addition. Hence, we need to first take the reciprocal of (13) and then compute the double integral. On the contrary, with the proposed SRFS-C3 model, we are able to obtain a closed form expression for average conductivity.

Before we conclude, it is worth mentioning that it may be possible to augment the circuit-compatibility of our model by formulating approximate analytical expressions for $\sigma_{SRFS}(0,0)$, $\sigma_{SRFS}(1,0)$, $\sigma_{SRFS}(0,1)$, $\sigma_{SRFS}(1,1)$ as function of $\kappa$ and $p$. Thus, by using the analytical model (instead of using the integral-based expressions of (1)) in conjunction with the SRFS-C3 approach, one can potentially speed up the computations. This can be investigated in a future work.

## VII. Conclusion

We proposed a 2D spatially resolved FS-based circuit-compatible conductivity (SRFS-C3) model for rectangular interconnects. The proposed model uses an analytical function to predict the spatial profile by using only four conductivity values from the physical SRFS model (derived in Part I [10]). This makes the SRFS-C3 model circuit compatible. Our model not only offers spatial resolution of conductivity (which matches closely with the physical SRFS model) but also accurately predicts the trends for a wide range of values of specularity, electron mean free path and interconnect dimensions. Additionally, we obtain a continuous closed form expression for the average conductivity using the SRFS-C3 model, which is valid for a wide range of physical parameters and overcomes the limitations of previous analytical models for average conductivity. We compare the SRFS-C3 model with a "cosh" based model [8] appropriately modified for a fair comparison. The key benefits of our approach are that (i) it is based on conductivity subtraction (which is more physical compared to resistivity addition that the modified "cosh" model uses) and (ii) it is integrable that enables an analytical formulation of the average conductivity (unlike "cosh" model) Overall, we find that the proposed SRFS-C3 model offers an excellent match to the SRFS model and is more accurate than the modified "cosh" model.